\documentclass[12pt,preprint]{aastex}
\begin{document}

\title{Jet Collimation by Small-Scale Magnetic Fields}

\author{Li-Xin Li}
\affil{Princeton University Observatory, Princeton, NJ 08544--1001, USA}
\email{E-mail: lxl@astro.princeton.edu}

\begin{abstract}
A popular model for jet collimation is associated with the presence of a
large-scale and predominantly toroidal magnetic field originating from 
the central engine (a star, a black hole, or an accretion disk). Besides 
the problem of how such a large-scale magnetic field is generated, in 
this model the jet suffers from the fatal long-wave mode kink 
magnetohydrodynamic instability. In this paper we explore an alternative 
model: jet collimation by small-scale magnetic fields. These magnetic 
fields are assumed to be local, chaotic, tangled, but are dominated by 
toroidal components. Just as in the case of a large-scale toroidal 
magnetic field, we show that the ``hoop stress'' of the tangled toroidal 
magnetic fields exerts an inward force which confines and collimates the 
jet. The magnetic ``hoop stress'' is balanced either by the gas pressure 
of the jet, or by the centrifugal force if the jet is spinning. Since 
the length-scale of the magnetic field is small ($<$ the cross-sectional 
radius of the jet $\ll$ the length of the jet), in this model the jet 
does not suffer from the long-wave mode kink instability. Many other
problems associated with the large-scale magnetic field are also 
eliminated or alleviated for small-scale magnetic fields. Though it 
remains an open question how to generate and maintain the required 
small-scale magnetic fields in a jet, the scenario of jet collimation 
by small-scale magnetic fields is favored by the current study on disk 
dynamo which indicates that small-scale magnetic fields are much easier 
to generate than large-scale magnetic fields.
\end{abstract}

\keywords{galaxies: jets --- stars: winds, outflows --- magnetic 
fields --- MHD}

~~
\newpage

%\section 1
\section{Introduction}
Jets are ubiquitous in astronomy. They exist in many astronomical
systems, e.g., quasars, galactic nuclei, stellar binaries, young stellar 
objects \citep{beg84,fer98,liv99,wii01}, and young pulsars \citep{got01}. 
Recently it has been suggested that jets also exist in gamma-ray bursts 
\citep[and references therein]{pir99,ber01}. A remarkable feature of jets 
is that they are highly collimated on a super-large length-scale compared 
to their cross-sectional radius \citep{beg84,fer98,liv99} except at the 
very beginning \citep{jun99,eis00}. Though more than eighty years have 
passed since the first jet was discovered in the center of the galaxy 
M87 \citep{cur18}, it is still a mystery how jets are generated and
collimated \citep{fer98,kro99,sik01,wii01}.

A popular model for jet collimation is associated with the existence of
a large-scale and predominantly toroidal magnetic field which originates 
from the central engine: a black hole, a star, or an accretion disk
\citep[and references therein]{ben78,bla82,beg84,fer98,kro99}. In this 
model, the jet is assumed to be launched from the central object [a star 
\citep[and references therein]{mes99}, or a black hole 
\citep{bla77,mac82,phi83}], or an accretion disk 
\citep{bla76,lov76,bla82,lov91,li92,fer93a,fer93b,fer95,shu94}.
An initially poloidal magnetic field threading the central object and the
accretion disk extracts angular momentum and energy from the central 
object and the accretion disk, which drives a magnetohydrodynamic (MHD)
outflow from the central object and the disk through the magnetosphere 
(corona) above them. At large radii, the inertia of plasma particles 
becomes important and bends the magnetic field lines so as to increase 
the toroidal component of the magnetic field. From the simple argument of 
the magnetic flux conservation, it is easy to show that at large radii 
the toroidal magnetic field always dominates the poloidal magnetic 
field \citep{beg84}. The dominant toroidal magnetic field produces an 
inward force (the ``hoop stress'') which confines the outflow into a 
collimated jet. In this model, the appearance of a super-large-scale 
magnetic field, which extends from the central engine into the body of 
the jet, is crucial for the jet collimation.

The association of jets with magnetized accretion disks or magnetized 
central objects (stars or black holes) is strongly supported by the 
recent observations of {\sl HST}, {\sl Chandra}, and {\sl VLA} (or {\sl 
VLBA}) on jets in galactic nuclei, protostellar systems, and young 
pulsars \citep{for94,har94,jun99,eis00,wei00,got01,gae01}. However,
though the scenario of jet collimation by a large-scale magnetic field
is simple and attractive, it has the following problems: (1) The origin 
of the large-scale magnetic field is unknown. It is unknown if such a 
large-scale magnetic field, which extends from the central engine into 
the body of the jet, could exist; and, if it exists, how the large-scale 
magnetic field is generated \citep{liv99}. In contrast, small-scale, 
chaotic, and tangled magnetic fields seem to be easily generated by 
dynamo processes in an accretion disk 
\citep[and references therein]{gal79,tou92,bal98,mil00}. Though some
effort has been made \citep{tou96}, it is still unclear if large-scale
magnetic fields can be generated from small-scale magnetic fields. 
(2) In this scenario, the jet is essentially continuously
accelerated and collimated by the central engine, since the large-scale
magnetic field is assumed to eventually connect to the central engine. But 
it is hard to understand how the central engine can continue controlling 
the dynamics of the jet far beyond the fast magneto-sonic point where the 
jet loses its causal connection to the central engine \citep{beg95,pac00}. 
(3) A large-scale magnetic field dominated by toroidal components is 
globally unstable \citep{bat78,fre87}. In particular, for a cylindrically 
shaped magnetic field with both poloidal and toroidal components, the 
long-wave mode kink instability (the so-called screw instability) sets in 
when the Kruskal-Shafranov criterion is satisfied \citep{kad66,bat78,fre87}
\begin{eqnarray}
    {2\pi R B_{\parallel} \over L B_{\perp}} < 1 \,,
    \label{ks}
\end{eqnarray}
where $L$ is the length of the cylinder, $R$ is the radius of the 
cylinder, $B_{\parallel}$ is the poloidal component of the magnetic field
which is parallel to the axis of the cylinder, and $B_{\perp}$ is the
toroidal component of the magnetic field which is perpendicular to the
axis of the cylinder. Equation (\ref{ks}) is equivalent to the statement
that the magnetic field is unstable when the magnetic field lines make
more than one turn about the axis over the cylinder. For a jet we have
$L \gg R$, equation (\ref{ks}) implies that the magnetic field is 
unstable when $B_{\perp} >B_{\parallel}$. Thus, a jet collimated by a 
large-scale toroidal magnetic field (which implies $B_{\perp} >
B_{\parallel}$) suffers from the long-wave mode kink instability and thus 
its global structure is expected to be twisted and disrupted quickly 
\citep{eic93,spr97,beg98,mes99,li00a,ker00}. 

Jet collimation by a large-scale poloidal magnetic field has also been 
proposed \citep{bla94,spr94,spr96,spr97}, but that raises the question 
of what collimates the poloidal magnetic field.

In this paper, we explore an alternative scenario: jet collimation by
small-scale magnetic fields. In this scenario, we assume the jet is 
threaded by small-scale, chaotic, and tangled magnetic fields which are
dominated by toroidal components. The length-scale of the magnetic field
lines is assumed to be $<$ the cross-sectional radius of the jet $\ll$
the length of the jet. This scenario is motivated by the investigation on 
disk dynamo which shows that chaotic and tangled magnetic fields 
dominated by toroidal components, and having typical length-scales
equal to the disk thickness, are efficiently generated in the disk
\citep{gal79,tou92,bal98,mil00}. Bubbles of magnetic fields may be driven 
into the disk corona by magnetic buoyancy \citep{gal79,cha94,mil00}, 
which may eventually form a MHD outflow. The dominance of the toroidal
components of the magnetic field in the disk is due to the differential 
rotation of the disk. If in the outflow the magnetic fields are still 
dominated by toroidal components --- the recent numerical simulation 
\citep{mil00} seems to indicate that this is true, then the outflow can 
be confined and collimated into a jet by the ``hoop stress'' of the
toroidal magnetic fields, just as in the case of a large-scale magnetic 
field. But the problems that appear in the case of a large-scale
magnetic field are eliminated or alleviated due to the small scale 
concerned in this scenario.

%\section 2
\section{Jet Collimation by Small-Scale Magnetic Fields}
Motivated by the disk dynamo investigation, we assume the jet is threaded
by small-scale, chaotic, and tangled magnetic fields. The magnetic fields
are dominated by toroidal components. The typical length of the magnetic
field lines is $l < R \ll L$, where $R$ is the cross-sectional radius
of the jet, $L$ is the length of the jet. This scenario naturally arises
if the jet is launched from a magnetized disk or a magnetized star. Though 
it is unclear if such a scenario can exist for a jet originating from a 
central black hole, with magnetic connection to a disk a rapidly rotating  
black hole may power a jet launched from the disk through pumping energy 
and angular momentum into the disk 
\citep[and references therein]{bla98,bla00,li00b,li00c}.
\citet{beg95} and \citet{hei00} have presented a model which shows how 
jets may be accelerated by tangled magnetic fields. As in their model, 
we assume that in the jet
\begin{eqnarray}
    \langle B_r\rangle = \langle B_\phi\rangle = \langle B_z
         \rangle = \langle B_r B_\phi\rangle = \langle B_\phi 
	 B_z\rangle = \langle B_z B_r\rangle =0 \,,
    \label{ass1}
\end{eqnarray}
but 
\begin{eqnarray}
    \langle B_r^2\rangle \neq 0 \,, \hspace{1cm}
    \langle B_\phi^2\rangle \neq 0 \,, \hspace{1cm}
    \langle B_z^2\rangle \neq 0 \,,
    \label{ass2}
\end{eqnarray}
where $r$, $\phi$, and $z$ are cylindrical coordinates with the $z$-axis
oriented along the jet axis, $B_r$, $B_\phi$, and $B_z$ are the components
of the magnetic field in the $r$, $\phi$, and $z$ directions, 
respectively, $\langle\rangle$ denotes suitable statistical average over 
space and time. In addition to their assumptions, we assume that
\begin{eqnarray}
    \langle B_\phi^2\rangle \gg \langle B_r^2\rangle \,, \hspace{1cm}
    \langle B_\phi^2\rangle \gg \langle B_z^2\rangle \,,
    \label{ass3} 
\end{eqnarray}
which assure that the magnetic fields are dominated by toroidal 
components. Unlike \citet{beg95} and \citet{hei00}, here we work in an
inertial frame (instantly) comoving with the motion of the jet along its 
axis. The cylindrical coordinates and the magnetic fields are defined
in such an inertial frame.

The special relativistic momentum equation for an ideal MHD fluid in an 
inertial frame is given by \citet{beg98}. If we choose the inertial frame 
to be (instantly) comoving with the translation motion of the jet along 
its axis, the momentum equation can be written as
\begin{eqnarray}
    \Gamma^2\left(\rho+{p\over c^2}\right)\left({\partial {\bf v}
         \over\partial t} + {\bf v}\cdot \nabla{\bf v}\right)
         = -\left(\nabla p + {{\bf v}\over c^2}{\partial p\over
           \partial t}\right)+ {1\over 4\pi}\langle\left(\nabla
           \times {\bf B}\right) \times {\bf B}\rangle \,,
    \label{mhd}
\end{eqnarray}
where ${\bf v}$ is the flow velocity, $\Gamma = \left(1 - v^2/c^2
\right)^{-1/2}$ is the Lorentz factor, $p$ and $\rho$ are the proper 
pressure and the proper mass density, respectively, ${\bf B}$ is the 
magnetic field, $c$ is the speed of light\footnote{For a perfectly 
conducting fluid the electric field in the comoving frame (i.e., the rest 
frame of the fluid) is vanishing. Through the Lorentz transformation, an 
electric field is induced in a non-comoving inertial frame, which is 
related to the magnetic field in the same inertial frame by ${\bf E} = 
-{{\bf v}\over c}\times {\bf B}$ where ${\bf v}$ is the velocity of the 
fluid in the inertial frame. The electric field induces a
Goldreich-Julian charge density in the inertial frame through the Gauss
law \citep{gol69}: $\rho_e = {1\over 4\pi} \nabla\cdot {\bf E}\,$. The
magnetic field induces a current density in the inertial frame through
the Ampere law including the displacement current: ${\bf j} = {c\over 4
\pi} \left(\nabla\times {\bf B} - {1\over c}{\partial\over \partial t}
{\bf E}\right)$. For relativistic motion ($v\approx c$), the electric
field may be comparable to the magnetic field, but can never dominate
the magnetic field since ${\bf B}^2 - {\bf E}^2$ is frame-independent and
in the comoving frame the electric field is vanishing. However, here
we choose the inertial frame to be comoving with the translation
motion of the jet along its axis, and in this inertial frame the only
possible relativistic motion of the jet is the rotation around its axis.
And, in this inertial frame, we assume the magnetic field is
predominantly toroidal [eq. (\ref{ass3})], which means that ${\bf v}$ is
almost parallel to ${\bf B}$. Thus, in the frame comoving with the
axi-translation motion of the jet, we always have $\langle E^2 \rangle
\ll \langle B^2\rangle$. Therefore, in equation (\ref{mhd}) we have
ignored the electric forces.\\
\indent\indent 
If the translation velocity of the jet along its axis varies with radius,
we choose the inertial frame to comove with the translation velocity at a 
radius $r$. Then equation (\ref{mhd}) applies to the neighborhood of $r$ 
in such an inertial frame.}. All quantities are defined in
the inertial frame, except that the pressure and the mass density are 
defined in the frame comoving with the fluid which is different from the
inertial frame if the jet rotates or expands radially. Since the magnetic 
field discussed in this paper is assumed to be
chaotic, tangled, and have small scale, in equation (\ref{mhd}) we average
the term $\left(\nabla\times {\bf B}\right) \times {\bf B}$ over space
and time. The scale over which the average is made is chosen to be
larger than the scale of the micro-structure of the magnetic field but
smaller than the macro-scale over which observations are taken.

Assuming the jet is approximately symmetric about the rotation around the
jet axis and the translation along the jet axis, then the average 
gradients of the magnetic field in the $\phi$ and $z$ directions are 
small and can be neglected. Then, with the assumptions in equations 
(\ref{ass1}--\ref{ass3}), we have
\begin{eqnarray}
    \langle\left(\nabla\times {\bf B}\right) \times {\bf B}\rangle
        &=& \langle{\bf B}\cdot\nabla{\bf B}\rangle - {1\over 2}
          \nabla \langle B^2\rangle \nonumber\\
        &\approx& -\hat{\bf r}\left({\langle B_\phi^2\rangle \over r}
          + {1\over 2}{\partial\over\partial r}\langle B_\phi^2
          \rangle\right) \nonumber\\
        &=& -{\hat{\bf r}\over 2 r^2} {\partial\over\partial r}
          \left(r^2 \langle B_\phi^2\rangle\right) \,,
    \label{dbb}
\end{eqnarray}
where $\hat{\bf r} \equiv {\bf r}/r$. In the right hand side of the 
second line of equation (\ref{dbb}), the first term represents the ``hoop 
stress'' which points inward toward the jet axis, and is nonzero even when 
the gradient of the magnetic pressure (the second term) is zero. From 
equation (\ref{dbb}), the jet can be self-collimated only if
\begin{eqnarray}
    {\partial\over\partial r}\left(r^2 \langle
        B_\phi^2\rangle\right) > 0 \,.
    \label{jetcol}
\end{eqnarray}

Since $\langle B_\phi^2\rangle = 0$ at the jet axis, equation 
(\ref{jetcol}) must hold close to the jet axis. At large radii,
$\langle B_\phi^2\rangle$ decreases with $r$, equation (\ref{jetcol})
is violated if $\langle B_\phi^2\rangle$ decreases faster than $r^{-2}$.
Then, if $\langle B_\phi^2\rangle$ decreases faster than $r^{-2}$ at large
radii, there must exist a radius $r_0$ where
\begin{eqnarray}
    \left.{\partial\over\partial r}\left(r^2 \langle
        B_\phi^2\rangle\right)\right\vert_{r=r_0} = 0 \,.
    \label{jetsuf}
\end{eqnarray}
Jet collimation can only happen in the region $r<r_0$. For $r>r_0$, 
the gradient of the magnetic pressure must dominate the ``hoop stress'', 
the flow cannot be confined by the magnetic ``hoop stress''. Thus, outside 
$r=r_0$ the flow must expand radially and eventually merge into the
surrounding medium where the magnetic pressure is balanced by the gas 
pressure of the surrounding medium. This indicates that around the 
collimated jet there exists an uncollimated thin ``corona'' which has a 
high ratio of the magnetic pressure to the gas pressure. This scenario 
is consistent with the speculation of \citet{li96} of the coexistence of 
a wide-angle component surrounding the well-collimated jet in young
stellar objects.

%\section 3
\section{A Toy Model: a Self-Collimated Jet Supported by the Centrifugal 
Force}
In this section, we consider a jet which has $p \ll \rho c^2$. For
simplicity, we assume the jet is steady, non-expanding, axisymmetric, 
and translation-symmetric along the jet axis. Then, in the inertial 
frame comoving with the jet motion along its axis, we have ${\bf v}
= \Omega r \hat{\phi}$ and ${\bf v}\cdot \nabla {\bf v} = -\Omega^2 r 
\hat{\bf r}$, where $\hat{\phi}$ is the azimuthal unit vector, $\Omega = 
\Omega (r)$ is the spinning angular velocity of the jet\footnote{We expect
the jet to spin since the jet can carry angular momentum from the central 
engine.}. Inserting equation (\ref{dbb}) into equation (\ref{mhd}), we 
obtain a one-dimensional equation
\begin{eqnarray}
    \Gamma^2 \rho \Omega^2 r \approx {dp \over dr} +{1 \over 8\pi r^2} 
        {d \over d r}\left(r^2 \langle B_\phi^2 \rangle\right) \,,
    \label{bal1}
\end{eqnarray}
which describes the dynamical equilibrium of the jet in the radial
direction.

In a thin shell close to the surface of the jet ($R-\delta < r <R$, 
$\delta\ll R$) we assume ${d\over dr} \langle B_\phi^2 \rangle \approx 
0$ and the jet is cold (thus $p\approx 0$). Then, in this surface shell, 
the jet is supported by the centrifugal force\footnote{Due to the complex
MHD turbulence, in a real jet thermal dissipation must happen so the gas
pressure is not negligible. Thus, a real jet does not seem to be 
supported purely by the centrifugal force. The toy model presented here
is mainly used to show, as the simplest example, how the jet collimation 
by small-scale magnetic fields works in principle.}, and from equation 
(\ref{bal1}) we have
\begin{eqnarray}
    \Gamma^2 \Omega^2 R^2 \approx {\langle B_\phi^2 \rangle \over
                            4\pi\rho} \,,
    \label{equ1}
\end{eqnarray}
where $\Gamma^2 = \left(1 - \Omega^2 R^2 /c^2\right)^{-1}$. In the 
non-relativistic limit, equation (\ref{equ1}) corresponds to an 
equipartition between the rotational energy and the magnetic energy.
One can check that the equilibrium between the centrifugal force and the
magnetic ``hoop stress'' is stable: if the radius $R$ increases a little 
bit, the ``hoop stress'' becomes greater than the centrifugal force,
making $R$ decrease and return to the original value in the equilibrium 
state; on the other hand, if $R$ decreases a little bit, the ``hoop 
stress'' becomes smaller than the centrifugal force, making $R$ increase 
and return to the original value in the equilibrium state.

We can express equation (\ref{equ1}) with conserved quantities. The rest
mass of the jet per unit length is $M_0 \approx \pi\rho R^2$. The 
angular momentum of the jet per unit length is $J_\phi\approx {1\over 2}
M\Omega R^2$, where $M c^2$ is the total energy of the jet per unit
length which is also a conserved quantity. If the jet is perfectly
conducting as we have assumed, the magnetic field lines are frozen in
the jet flow. Though for a chaotic magnetic field the macroscopic net 
magnetic flux through a macroscopic surface
is always zero, we can define a conserved absolute magnetic flux by
$\Psi \equiv\int \vert {\bf B}\cdot d{\bf S}\vert$, where $d{\bf S}$ is
the area element. The absolute magnetic flux $\Psi$ represents the total
number of magnetic field lines passing through a surface with an area
$S$, regardless of the direction of the magnetic field lines. For the
toroidal magnetic field, we have $\Psi_\phi \approx \langle B_\phi^2
\rangle^{1/2} R$ for per unit length of the jet. Inserting $M_0$, 
$J_\phi$, and $\Psi_\phi$ into equation(\ref{equ1}), we obtain
\begin{eqnarray}
    R \approx {2 J_\phi \over Mc} \left(1 + {4 M_0 c^2 \over
        \Psi_\phi^2}\right)^{1/2} \,,
    \label{radius}
\end{eqnarray}
and
\begin{eqnarray}
    \Omega \approx {Mc^2\over 2 J_\phi} \left(1 + {4 M_0 c^2 \over
        \Psi_\phi^2}\right)^{-1} \,.
    \label{omega}
\end{eqnarray}
The spin velocity on the surface of the jet is
\begin{eqnarray}
    V = \Omega R \approx {c\over\sqrt{1 + {4 M_0 c^2 \over
        \Psi_\phi^2}}}\,,
\end{eqnarray}
which as expected can approach but cannot exceed the speed of light.

In the non-relativistic limit (i.e., $\Psi_\phi^2 \ll M_0 c^2$), we have 
$M\approx M_0$, $R\approx 4 J_\phi / \left(M_0^{1/2}\Psi_\phi\right)$, 
$\Omega \approx \Psi_\phi^2/\left(8 J_\phi\right)$, and $V\approx
\Psi_\phi/\left(2 M_0^{1/2}\right)$. In the relativistic limit (i.e., 
$\Psi_\phi^2 \gg M_0 c^2$), we have $R\approx 2J_\phi/\left(Mc\right)$, 
$\Omega\approx Mc^2/\left(2J_\phi\right)$, and $V\approx c$.

%\section 4
\section{Discussion and Conclusions}
As in the case of a large-scale and predominantly toroidal magnetic 
field, small-scale and predominantly toroidal magnetic fields can 
produce a ``hoop stress'' which confines and collimates the jet. Compared 
to the scenario that jets are collimated by a large-scale magnetic field, 
the scenario that jets are collimated by small-scale magnetic fields has 
the following
features: (1) Small-scale magnetic fields are likely to be easier to
create through disk dynamo processes \citep{gal79,tou92,bal98,mil00}. 
(2) Once the outflow of small-scale magnetic bubbles leaves the central 
engine (e.g., a disk), the outflow is disconnected from the central 
engine. The jet formed from such outflows is completely 
self-collimated --- in the sense that the small-scale magnetic fields 
in the jet do not connect to the central engine so the dynamics of the 
jet is not controlled by the central engine. (3) Since the magnetic 
fields are assumed to exist only on small-scales, any MHD instability 
would happen only on a length-scale $<$ the radius of the 
jet $\ll$ the length of the jet. Such small-scale MHD instabilities are 
not expected to destroy the global structure of the jet unless the 
relevant dissipation and evolution processes are so strong that the 
small-scale toroidal magnetic fields are eliminated so that the 
conditions in equation (\ref{ass3}) are violated. In other words, a jet 
collimated by small-scale magnetic fields is immune to the large-scale 
MHD instability.

Almost all models of disk dynamo predict that the chaotic and tangled 
small-scale magnetic fields produced in the disk are dominated by 
toroidal components because of the differential rotation of the disk. 
Thus, if a jet originates from such a disk, at the beginning of the jet 
the magnetic fields are already dominated by toroidal components. In
the disk the centrifugal force is balanced by the gravity of the central
object (a black hole or a star). As bubbles of small-scale magnetic
fields leave the disk, the gravity decreases. So it is reasonable to 
assume that at the beginning of the jet the centrifugal force prevails
and makes the outflow expand radially. Due to the conservation of the
absolute magnetic flux which is defined in the last section, we have 
$\langle B_r^2 \rangle \propto r^{-2}$, $\langle B_\phi^2\rangle \propto 
r^{-2}$, and $\langle B_z^2 \rangle\propto r^{-4}$ as $r$ --- the radius 
of the flow --- increases. Therefore, the expansion of the outflow 
increases the ratio $\langle B_\phi^2 \rangle / \langle B_z^2\rangle$ but 
keeps the ratio $\langle B_\phi^2 \rangle / \langle B_r^2\rangle$. As $r$ 
increases, the density of the rotational energy decreases as $\rho 
\Omega^2 r^2 \propto r^{-4}$, since $\rho\propto r^{-2}$ due to the
conservation of mass, and $\Omega\propto r^{-2}$ due to the conservation
of angular momentum. Thus, the radius will increase to a value $R$ when 
the magnetic ``hoop stress'' balances the centrifugal force (and the 
pressure gradient if the jet is not cold), i.e. when equation (\ref{equ1}) 
is satisfied. Then the jet will stop expanding and be collimated at the 
radius $R$ since the equilibrium between the magnetic ``hoop stress'' and
the centrifugal force is stable. In this picture, the magnetic 
field in the jet is dominated by the toroidal components at the birth of
the jet, and this feature is kept at large distance in the jet if the 
magnetic flux is conserved. However, we point out that even if initially
in the jet flow the toroidal magnetic field is comparable to the 
poloidal magnetic field, at large radii we also expect $\langle B_\phi^2
\rangle > \langle B_z^2\rangle$ if the magnetic flux is conserved, just
as in the case of a large-scale magnetic field.

In this paper we have not addressed the problem how jets are produced
and accelerated. With three-dimensional MHD simulations, 
\citet{mil00} have shown that the dynamo process associated with the 
MHD turbulence driven by the magneto-rotational instability (MRI)
\citep{bal91,bal98} works very effectively in a weakly magnetized 
accretion disk. About $25 \%$ of the magnetic energy generated by the 
MRI in the disk escapes due to buoyancy, producing a strongly magnetized 
corona above the disk. They have shown that in both the disk and the 
corona the magnetic fields are dominated by toroidal components. However,
due to the limits of the length-scales in the simulation, \citet{mil00} 
have not addressed the problem if a MHD outflow is produced. If a MHD 
outflow is produced from such a disk and corona, we expect that in the 
outflow the magnetic fields are chaotic, tangled, dominated by toroidal 
components, and have small length-scales. As the outflow gets far from 
the black hole, the magnetic ``hoop stress'' becomes dominant in the 
flow dynamics, confines and collimates the outflow into a jet.
We have also not considered the dissipation and evolution of the magnetic
fields in jets. Small-scale and tangled magnetic fields tend to reverse 
signs frequently, thus magnetic reconnection must happen; this will 
annihilate magnetic fields, and lead to energy dissipation. Jets are 
likely to spin differentially. Due to the differential spin and the large 
velocity gradient near the surface of the jet, small-scale shear 
instabilities tend to generate ripples and vortices as shown by numerical 
simulations \citep{mat90a,mat90b,cer01}, which will make the structure 
of the jet much more disordered than assumed here. Furthermore, the shear 
motion introduced by the large velocity gradient near the surface of the 
jet will generate strong poloidal magnetic fields which tend to weaken 
the self-confinement of the jet \citep{beg84,mat90a,mat90b,ker00}. These 
processes are important for the dissipation and evolution of the magnetic 
fields in jets, but it is unclear if they are strong enough to destroy the 
collimation mechanism provided by the small-scale magnetic fields.

We note that \citet{beg95} and \citet{hei00} have considered the problem 
of jet acceleration by tangled magnetic fields, but in their model the 
jet is assumed to be collimated by the thermal pressure from an external 
medium. \citet{beg95} has argued that, since tangled magnetic fields 
include large pressure gradients as well as tension forces, it is 
doubtable if a local equilibrium can be reached. However, we point out 
that, since we assume the length-scales of the magnetic fields are small, 
all large pressure gradients and tension forces exist only on small scales
which are comparable to the length of the field lines. On macroscopic
scales, the average pressure gradients and tension forces need not be
large. To see this, let us consider a simple example: $B_r = 0$, $B_\phi
= A (r/r_1) f(r) \sin (m \phi) \cos (k z)$, $B_z = f(r) \cos (m \phi) 
\sin (k z)$, where $A$, $r_1$, $m$, and $k$ are constants. Since ${\bf B}
(\phi+2\pi) = {\bf B}(\phi)$, $m$ must be an integer. To satisfy $\nabla
\cdot {\bf B} = 0$, we must have $k = -m A / r_1$.  Assuming $m$ and $k$ 
are large, then $B_\phi$ and $B_z$ reverse signs frequently. 
If we choose the length-scale on which the average is taken to be 
larger than the periods of the magnetic field, then we have $\langle 
B_\phi\rangle = \langle B_z\rangle = \langle B_\phi B_z\rangle = 0$, 
$\langle B_\phi^2\rangle = A^2 (r/r_1)^2 (f^2 /4)$, and $\langle B_z^2
\rangle = f^2 /4$. Then, if $r > r_1$ and $A^2 \gg 1$, we have $\langle
B_\phi^2\rangle \gg \langle B_z^2\rangle$. Since $k$ is large, $d 
B_\phi^2 / dz \propto k$ is also large. However, we have $\langle 
d B_\phi^2 / dz\rangle = d\langle B_\phi^2\rangle /dz \propto \langle 
\sin (k z) \cos (k z)\rangle = 0$. This is because within the macroscopic 
average scale the pressure gradients change signs rapidly thus they 
cancel each other. Certainly the local fluctuation of magnetic fields must
be accompanied by the corresponding local fluctuation of gases, so that 
the force balance is maintained. \citet{beg95} has also argued that the 
MHD turbulent motion in the jet tends to make the overall stress less 
anisotropic which will in turn weaken the self-collimation of the jet, 
but how important this effect will be remains to be shown by numerical 
simulations. Tangled magnetic fields in jets have also been discussed by
\citet{lai80,lai81}, who claims that the emission of highly polarized
synchrotron radiation by a source does not necessarily imply a highly 
ordered magnetic field. \citet{lai80} has also shown how the shear motion
in jets can compress an initially random field into an anisotropic 
configuration.

In summary, we have presented an interesting model for jet collimation 
alternative to the popular model with a large-scale magnetic field. We 
have shown that small-scale and predominantly toroidal magnetic fields 
are promising for confining and collimating jets. Almost all the problems 
associated with the case of a large-scale magnetic field are eliminated or 
alleviated for the case of small-scale magnetic fields, though many other 
problems remain to be addressed, e.g. the production and the acceleration 
of jets by small-scale magnetic fields, and the dissipation and evolution 
of small-scale magnetic fields in jets.

\acknowledgments
I am very grateful to Bohdan Paczy\'nski for many stimulating discussions, 
to Mitchell Begelman and Paul Wiita for many valuable comments. This work 
was supported by the NASA grant NAG5-7016 and a Harold W. Dodds Fellowship 
of Princeton University.

\newpage

%REFERENCES

\end{document}